\documentstyle[aps,prl,twocolumn,floats,epsfig,times]{revtex}
\begin{document}

\draft

\def\overlay#1#2{\setbox0=\hbox{#1}\setbox1=\hbox to \wd0{\hss #2\hss}#1%
  \hskip -2\wd0\copy1} \twocolumn[
\hsize\textwidth\columnwidth\hsize\csname@twocolumnfalse\endcsname

\title{Ultrahigh reflection from a medium with ultraslow group velocity}

\author{V.V. Kozlov, S. Wallentowitz, and S. Raghavan}

\address{Department of Physics and Astronomy and Rochester Theory
  Center for Optical Science and Engineering \\ University of
  Rochester, Rochester, New York 14627}

\date{\today}

\maketitle

\begin{abstract}
  We show that an incident wavepacket at the boundary to a medium with
  extremely slow group velocity, experiences enhanced reflection and a
  substantial spatial and temporal distortion of the transmitted wave
  packet. In the limit of vanishing group velocity, light cannot be
  transferred into the medium due to its perfect reflectivity.
\end{abstract}

\vskip2pc]


Currently there is a great deal of interest in slowing down light in
dispersive media. It is triggered by a continuous experimental
progress using electromagnetically induced transparency (EIT) in
gaseous media. The attained group velocities are as small as 17 m/s
for the medium being a Bose--Einstein condensate, 90 m/s in a hot
rubidium gas, and even 8 m/s in a rubidium vapor at room
temperature~\cite{slow}. EIT provides here a large enhancement of the
dispersion of medium while absorption is largely
suppressed~\cite{eit}.

A theoretical description of optical radiation propagating with group
velocities $v_g \!\sim\! 10^{-8} c$ (where $c$ is the speed of light
in vacuum) or even slower, may require a careful reconsideration of
some approximations routinely employed in optics. For example, the
reflection of an optical beam at the boundary of the medium is usually
considered to be determined by the refractive index of the medium.
This is perfectly valid for a monochromatic beam, and serves as a
perfect approximation for a wavepacket in a typical dielectric with
weak dispersion, where each spectral component experiences essentially
the same refractive index. However, when the dispersion steepens, a
substantial variation of the refractive index over the spectrum of a
wavepacket may be present. Then the usual approximation fails and new
effects emerge that are dominated by the spectrally dependent
refraction of the transmitted light.  Besides a severe spatio-temporal
distortion of the transmitted beam, one might expect difficulties in
actually transferring the incident optical wavepacket into a medium of
such strong dispersion.

In this Letter we show the significance of these effects for optical
experiments with extremely slow group velocity. We point out a
fundamental obstacle preventing one from infinitely decreasing the
speed of light inside a medium: The slowing down of light inside the
medium is accompanied by an increase of the reflectivity of the
boundary of the medium. Near the limit of stopping light, almost no
energy can penetrate from outside (as well as leak from inside) and
the boundary of the medium will behave as a perfect mirror.

Let us consider the reflection and transmission of an optical
wavepacket at a plane boundary between vacuum and a linear dielectric
with frequency dependent dielectric function $\varepsilon(\omega)$.
Since a relatively small absorption will have little effect on our
conclusions we take $\varepsilon(\omega)$ to be real
valued~\cite{remark}.  Expanding the dielectric function of the medium
around the center frequency $\omega_0$ of the incident wavepacket and
keeping only the leading term, we arrive at
\begin{equation}
  \label{eq:dispersion}
  \varepsilon(\omega) = n_0^2 + \frac{2\alpha}{\omega_0} \, (\omega
  \!-\! \omega_0) + \ldots , 
\end{equation}
where $n_0$ is the (absolute) refractive index of the medium at
$\omega \!=\! \omega_0$ and $\alpha$ determines the steepness of the
dispersion~\cite{remark2}. In the medium, the magnitude of the wave
vector and the dielectric function are related by $k(\omega) \!=\!
\omega / c \sqrt{\varepsilon(\omega)}$, from which we obtain for the
group velocity inside the medium,
\begin{equation}
  \label{eq:group-velocity}
  v_g = \frac{c}{n_0 \!+\! \alpha / n_0} .
\end{equation}

\begin{figure}[h]
  \begin{center}
    \epsfig{file=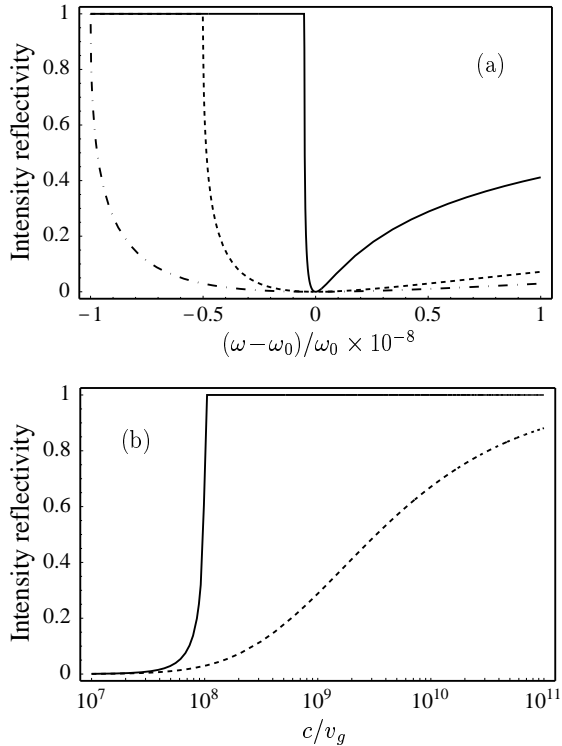,scale=0.55}
  \end{center}
  \caption{Dependence of $R(\omega)$ on (a) the scaled frequency
    $(\omega \!-\! \omega_0)/\omega_0$ for group velocities $c/v_g$:
    $5 \!\times\!  10^7$ (dot-dashed), $10^8$ (dashed), and $10^9$
    (solid), and (b) on the group velocity $v_g$ at the low and high
    frequency tails of a 1$\mu{\rm s}$ pulse, i.e. for $(\omega \!-\!
    \omega_0)/\omega_0$: $-5 \!\times\! 10^{-9}$ (solid) and $5
    \!\times\! 10^{-9}$ (dashed).}
  \label{fig:reflection}
\end{figure}

The transmittance at the boundary of the medium can be illustrated in
more detail by considering the intensity reflection and transmission
coefficients for the case of normal incidence~\cite{jackson}:
\begin{equation}
  R (\omega ) = \left| \frac{1 \!-\! \sqrt{\varepsilon(\omega)}}
    {1 \!+\! \sqrt{\varepsilon(\omega)}} \right|^2 , \quad
  T (\omega ) = \left| \frac{2}{1 \!+\! \sqrt{\varepsilon(\omega)}}
  \right|^2 , 
\end{equation}
As can be seen from Fig.~\ref{fig:reflection}, slowing down the group
velocity brings forth two distinct features in the reflectivity: (a)
For frequencies below the increasing cutoff frequency,
\begin{equation}
  \label{eq:cutoff}
  \omega_c = \omega_0 \left( 1 \!-\! \frac{n_0^2}{2\alpha} \right) ,   
\end{equation}
the low-frequency wing experiences total reflection, and (b) an
overall increase of the reflectivity for the high-frequency wing.
Usually these effects play no role in optical experiments. However,
for a small ratio $v_g/c$, the cutoff frequency may move into the
spectrum of the incident optical wavepacket. One clearly observes from
Fig.~\ref{fig:reflection}, that whereas for $c/v_g \!=\! 5 \!\times\!
10^7$ and a wavepacket of 1$\mu{\rm s}$ duration the reflectivity is
still small, for lower group velocities the cutoff frequency
approaches $\omega_0$, while the reflectivity for the high-frequency
tail also continuously increases. In the extreme limit $v_g/c \!\to\!
0$ we then obtain overall reflection of the incident optical
wavepacket for all frequencies apart from the center frequency.
%

The parameter $r$ defined as the ratio of the characteristic spectral
width of the incident wavepacket $\Delta\omega$ and the passband,
\begin{equation}
  r = \frac{\Delta\omega/2}{\omega_0 \!-\! \omega_c}
  = \frac{\alpha}{n_0^2} \, \frac{\Delta \omega}{\omega_0} 
  \approx \frac{c/n_0}{v_g} \, \frac{\Delta \omega}{\omega_0} ,
  \label{eq:parameter}
\end{equation}
can be used as a measure for the relevance of the boundary effects.
When $r \!>\! 1$ the cutoff frequency considerably moves into the
spectrum of the incident wavepacket, so that the variation of the
refractive index over the spectrum leads already to an enhanced
reflection and substantial spectral (temporal) distortion of the wave
packet. The spectral modifications are asymmetric with respect to the
center frequency $\omega_0$. Since the low-frequency tail experiences
a larger reflection compared to the higher frequencies, this results
in an overall blue shift of the transmitted wavepacket.

For the ideal case of EIT with purely radiative transitions of rate
$\gamma$, negligible Doppler effect, and long-living coherence, one
finds $n_0 \!\approx\! 1$ and $r \!\approx\! 3 / (8\pi^2) N\lambda^3
\gamma \Delta\omega / \Omega^2$ with $N$ and $\Omega$ being the atomic
density and the Rabi frequency of the driving field, respectively. The
relevant spectral width of the wavepacket may be taken as the EIT
transparency window $\Delta\omega \!\approx\! \Omega^2/\gamma$, and we
immediately obtain the estimation $r \!\approx\! 3 N \lambda^3 /
(8\pi^2)$.  In conclusion, for optical fields, say of wavelength
$\lambda \!=\! 800 \, {\rm nm}$, $r$ may already approach unity for
rather modest densities of $N \!\sim\! 5 \!\times\! 10^{13} {\rm
  cm}^{-3}$.

\begin{figure}
  \begin{center}
    \epsfig{file=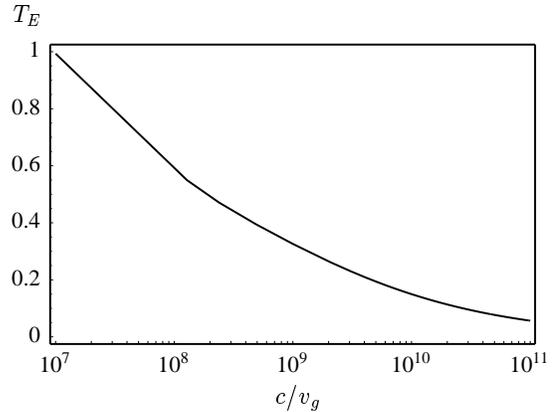,scale=0.55}
  \end{center}
  \caption{Dependence of $T_E \!=\! E_t/E_i$ on the group velocity
    $v_g$ for the center frequency and spectral width of the incident
    Gaussian wavepacket being $\omega_0 \!=\!  10^{14} {\rm Hz}$ and
    $\Delta\omega \!=\!  10^6 {\rm Hz}$.}
  \label{fig:energy-transfer}
\end{figure}

In general, the importance of the boundary effects can be demonstrated
by the energy transfer into the medium, characterized by the ratio:
$T_E \!=\! E_t / E_i$. Here $E_t$ and $E_i$ are the total energies of
the transmitted and incident wavepackets, respectively, defined as the
normal component of the Poynting vector in and outside the medium,
integrated over time and the transverse area of the beam.
Fig.~\ref{fig:energy-transfer} shows a typical example of how the
energy-transfer ratio $T_E$ changes with group velocity for an
incident Gaussian wavepacket. The continuous decrease of $T_E$ with
decreasing group velocity indicates a vanishing energy flow through
the boundary for $v_g/c \!\to\! 0$.
The above result remains conceptually valid for a dielectric function
of a more general form than Eq.~(\ref{eq:dispersion}). Given that any
additional term appearing in Eq.~(\ref{eq:dispersion}) is small
compared to the first-order term, we expect only a minor modification
of the cutoff frequency and of the overall index of refraction.

Analogous effects arise when the sharp boundary is replaced by a layer
of finite width $L$, where the dielectric function varies continuously
along the direction of propagation. Such a regime is readily available
in EIT with the continuous variation of $\varepsilon$ being provided
by the externally controllable driving field. Despite arbitrary
intermediate modifications, the reflection properties depend only on
the difference between the initial and final values of the dielectric
function, $\varepsilon (\omega, z\!=\! 0)$ and $\varepsilon (\omega,
z\!=\! L)$, respectively~\cite{born-wolf}. Then our results are
recovered upon replacement of $\varepsilon(\omega)$ by
$\varepsilon(\omega, z \!=\!  L)$ in Eq.~(\ref{eq:dispersion}).

For the more general situation where the incident beam has some
spread, we expect along with temporal also spatial distortions.
Typically the spread of the transmitted beam will increase with
increasing refractive index. Approaching the extreme case of very
large dispersion, when the cutoff frequency appears inside the
spectrum, the lower frequency components are totally reflected,
whereas the other components are refracted over a wide range of
angles.

The enhanced reflectivity from the highly dispersive medium draws
attention also to the importance of backscattering effects in the
field dynamics in inhomogeneous/nonlinear media. Whenever the
parameter $r$ (or $1/T_E$) gains appreciable values, the widely used
approximation of unidirectional propagation may fail. Instead, the
full wave equation must be used for a proper inclusion of the
counter-propagating waves.

\begin{figure}
  \begin{center}
    \epsfig{file=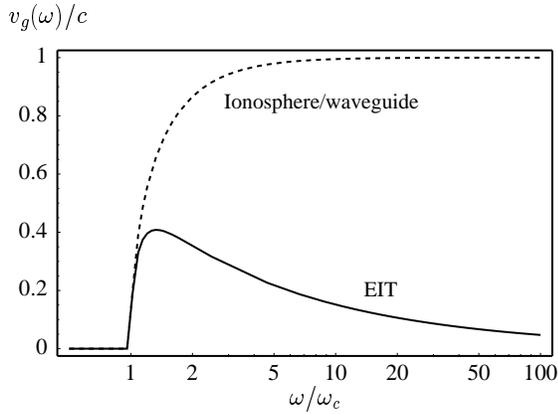,scale=0.55}
  \end{center}
  \caption{Comparison of the frequency dependence of group velocities
    for the case considered here (solid) and the case of the
    ionosphere or a waveguide (dashed). The solid curve has been
    amplified by the factor $10^4$ and corresponds to a group velocity
    at $\omega_0$ of $v_g/c \!=\! 10^{-8}$, for both curves $n_0 \!=\!
    1$.}
  \label{fig:comparison}
\end{figure}

The occurrence of a cutoff frequency where wave propagation is
inhibited is a well known phenomenon in the ionosphere and
waveguides~\cite{jackson}. However, compared to these cases we deal
here with a dielectric function of a conceptually different form. To
illustrate this, we compare in Fig.~\ref{fig:comparison} the frequency
dependence of the group velocity for these cases. Whereas, in the case
of the ionosphere or waveguides the group velocity is approaching $c$
for increasing frequency, in our case a maximum can be observed, after
which the group velocity is continously falling to zero. That is, in
our case a suppression of wave propagation is present also for higher
frequency.

In conclusion, an incident wavepacket at the boundary to a highly
dispersive medium with extremely slow group velocity experiences
enhanced reflection and a substantial spatio-temporal distortion of
the transmitted wavepacket. In the case of EIT, the frequency
selective reflectivity may be utilized for generating pulses (in
reflection) revealing spectral shapes sensitively controllable by the
external driving field. One may also expect that spectrally resolved
polarization measurements of the reflected beam can provide an
efficient tool for probing the dispersive properties of such media. On
the other hand, the transmission losses due to the enhanced
reflectivity pose a serious restriction on possible applications of
slow light for quantum-optical purposes. Finally, we have shown, that
in the extreme limit of vanishing group velocity, light cannot
penetrate the boundary due to the perfect reflectivity of the medium.

S.W. acknowledges support from the Studienstiftung des deutschen
Volkes by a BASF research fellowship.


\begin{thebibliography}{99}
\bibitem{remark} For a dispersive medium the imaginary and real parts
  of $\varepsilon(\omega)$ are linked via a Kramers--Kronig relation.
\bibitem{remark2} Without loss of generality, we consider here the
  case $\alpha \!\geq\! 0$. The opposite case is easily obtained by
  exchanging the low and high frequency tails of the wavepacket.
\end{thebibliography}
\end{document}